%Paper: cond-mat/9509165
%From: Anne Bohle <bohle@mpidx3.mpip-mainz.mpg.de>
%Date: Wed, 27 Sep 1995 09:37:32 +0100

\magnification=1200
\vsize=8.5truein
\hsize=6truein
\baselineskip=20pt
\centerline{\bf Polydispersity and ordered phases in solutions of
rodlike macromolecules}
\centerline{by}
\centerline{Anne Bohle$^1$, Robert Ho{\l}yst$^{2}$ and Thomas Vilgis$^1$}
\vskip 20pt
\centerline {$^1$ Max-Planck-Institut f\"ur Polymerforschung, D-55021 Mainz,
Postfach 3148, FRG}
\centerline {$^2$Institute of Physical Chemistry PAS and College of Sciences}
\centerline {Dept. III, Kasprzaka 44/52,
01224 Warsaw, Poland}
\centerline{\bf{Abstract}}
We apply density functional theory to study the influence of polydispersity on
the stability of columnar, smectic and solid ordering in the solutions of
rodlike macromolecules. For sufficiently
large length polydispersity (standard deviation $\sigma>0.25$)
a direct first-order
nematic-columnar transition is found, while for smaller $\sigma$ there is a
continuous nematic-smectic and first-order smectic-columnar transition.
For increasing polydispersity the columnar structure is stabilized
with respect to solid perturbations.
The length distribution of macromolecules changes neither
at the nematic-smectic nor at the nematic-columnar transition, but
it does change at the smectic-columnar phase transition.
We also study the phase behaviour of binary mixtures, in which
the nematic-smectic transition is again found to be continuous. Demixing
according to rod length in the smectic phase is always
preempted by transitions to solid or columnar ordering.

\vfill\eject

It has been known for some time that concentrated solutions of DNA$^1$,
po\-lypeptides$^{2,3}$, polysaccharides$^3$ and hairy rod
polymers$^4$ form columnar
phases. It also has been
observed that DNA in bacteriophages and sperm nuclei of
sepia, trout and salmon
exhibit columnar ordering$^1$.
The  activity of DNA  (renaturation, transcription or
replication) can be enhanced in the condensed  phase$^5$.
Also, the
condensed form of DNA can be used by nature to store genetic material
in small volume and use it at the moment of cell cycle.
Despite the accumulating body of experimental data,
very little is known about the influence of various factors, such as
attractive forces or polydispersity, on the stability of columnar
ordering in macromolecular solutions.
It is known that columnar ordering is preempted by smectic ordering in
hard monodisperse rod systems$^6$, but can be stabilized in binary
mixtures of
rods of different length$^7$. However, since in actual solutions the systems
are characterized by the continuous distribution of molecular length, true
monodisperse or bidisperse
systems are rare.
Here we fill this apparent gap and study, within Density
Functional Theory$^{8,9}$ (DFT),
the influence of polydispersity on columnar ordering.

A polydisperse system  with continuous distribution of molecular masses
(or length as is the case here) can be regarded as a mixture of infinitely
many components. Thus, phase equilibria between two phases requires
the equality of chemical
potentials for molecules of all lengths, making the total number of
equilibrium conditions infinite. Previous studies of polydisperse systems have
employed bifurcation analysis$^{10}$ or expansion in the distribution function
width$^{11}$ (valid for sharp distributions only), but, to date, no general
approach is known.
Here we establish the equilibrium conditions by a novel technique that does not
involve expansion or assumption of the sharpness of the molecular length
distribution.

We pose the following questions: What is the minimal degree of
polydispersity necessary to stabilize the columnar phase?
At what polydispersity is the smectic phase destabilized?
Does polydispersity change the continuous nature of the nematic-smectic
phase transition? Are there any smectic phases in which rods of different
lengths are demixed (completely demixed, or ordered in layers with
varying widths)?

We consider a polydisperse
system of hard, parallel, cylinders of diameter $D$ interacting via the
hard core repulsion potential.
The free energy of the system as a functional of the
number density $\rho_0({\bf r},L)$, for a given length $Ll$
($L$ is dimensionless), is given
by the following formula (the free energy is in $k_BT$ units):
$$F[\rho]=\int dL\int d{\bf r}\rho_0({\bf r},L)
\left(\ln{\left(\lambda^3\rho_0({\bf r},L)
\right)}-1\right)+\int dL\int d{\bf r}\rho_0({\bf r},L)\Psi\left
(v_0\bar\rho_0({\bf r},L)\right).\eqno(1)$$
where $\lambda$ is the De Broglie wavelength. The first term in Eq(1) is
exact and represents the configurational entropy of the polydisperse system.
The second term is the excess free energy
determined by the interparticle interactions. Here $\Psi$ is
the excess free energy density of the homogeneous system and
$\bar\rho_0({\bf r},L)$ is the weighted density, defined by
$$\bar\rho_0({\bf r},L)=\int dL'\int d{\bf r}'w({\bf r}-{\bf r}',L+L')
\rho_0({\bf r}',L'),\eqno(2)$$
where the weight function $w({\bf r}-{\bf r}',L+L')$ is normalized according to
$$\int d{\bf r}w({\bf r}-{\bf r}',L+L')=1\eqno(3)$$
The weighted density (Eq(2))
represents the influence of the total density of particles
on the density of rods of length $L$ at point ${\bf r}$.
The normalized length distribution of the
cylinders is
given by:
$$g(L)={1\over N}\int d{\bf r}\rho_0({\bf r},L),\eqno(4)$$
where $N$ is the number of particles in the system.

Now we make the following approximations. First, for the weight
function we choose the normalized Mayer function, i.e.:
$$\eqalign{w({\bf r}-{\bf r}',L+L')=
&{{\Theta\left((L+L')l/2-\vert z-z'\vert\right)}
\over{(L+L')l}}{{\Theta\left(D-\vert {\bf r}_\bot-{\bf r}_\bot'\vert\right)}
\over{\pi D^2}}\cr= &f_1(L+L',\vert z-z'\vert)f_2(D,\vert{\bf r}_\bot-
{\bf r}'_\bot\vert),\cr}\eqno(5)$$
where $\Theta$ is the Heaviside step function.
This weight function implies that the free energy of a system with columnar
ordering does not depend on the degree of polydispersity.
Second, we employ a decoupling approximation, i.e.,
$$ \rho_0({\bf r},L)=\rho({\bf r})g(L).\eqno(6)$$
For the columnar phase Eq(6) is actually exact.
For the smectic phase the approximation is expected to break
down only if the width of the layers varies. As we show below
(for binary mixtures) however, this does not happen.
The density distribution $\rho({\bf r})$ is approximated
by a Gaussian function centered at the sites of the
Bravais lattice characteristic for the given phase. It follows from the studies
of hard
spheres fluids$^{8,13-17}$ that this is a very good approximation.
We assume that $g(L)$ is given by the Gaussian distribution, characterized
by the mean value $L_0$, and the standard deviation $\sigma$. It turns out
that the mean length $L_0$ scales out.
Finally, the excess free energy density for the homogeneous system is
approximated by the Carnaham-Starling equation as in Ref.6, i.e.,
$$ \Psi(\eta)={{\eta (4-3\eta )}\over{(1-\eta )^2}},\eqno(7)$$
where $\eta=\rho v_0$, $\rho$ is the average number density and
$v_0=(1/4)\pi D^2lL_0$ is the mean volume of the cylinders.

The phase diagram obtained from the above outline is shown in Fig.1.
It encompasses the nematic, smectic, columnar and solid phases.
For $\sigma\ge 0.25$ there is a direct first order phase transition to the
columnar phase. The chemical potential as a function of $L$ for the nematic
and columnar phases can be written in the form
$$\mu(L)=\ln{(g(L))}+\mu_0,\eqno(8)$$
It is central to our approach that for at least one of the phases
in equilibrium $g(L)$ may be calculated as a function of $\mu(L)$, as is the
case for Eq(8). We make use of this in Eq(13).
In the nematic phase $\mu_0 $ is given by
$$\mu_0^{\rm nem}(L)=\ln{(\lambda^3\rho^{\rm nem})}+\Psi(v_0\rho^{\rm nem})+
v_0\rho^{\rm nem}\Psi'(v_0\rho^{\rm nem})\eqno(9)$$
and in the columnar phase by
$$\eqalign{&\mu_0^{\rm col}={{\displaystyle{\int d{\bf r}_\bot
\rho^{\rm col}({\bf r}_\bot)
\ln{(\lambda^3\rho^{\rm col}({\bf r}_\bot))}}}\over{\displaystyle{
\int d{\bf r}_\bot\rho^{\rm col}
({\bf r}_\bot)}}}+\cr
&{{\displaystyle{\int d{\bf r}_\bot
\left(\rho^{\rm col}({\bf r}_\bot)\Psi(v_0\bar\rho^{\rm col}({\bf r}_\bot))
+
v_0\rho^{\rm col}({\bf r}_\bot)
\bar\rho^{\rm col}({\bf r}_\bot)\Psi'
(v_0\bar\rho^{\rm col}({\bf r}_\bot))\right)}}
\over{\displaystyle{\int d{\bf r}_\bot\rho^{\rm col}
({\bf r}_\bot)}}}\cr}
\eqno(10)
$$
The equilibrium density distribution in the columnar phase
$\rho^{\rm col}({\bf r}_\bot)$ is a sum of Gaussian functions centered
at the sites of a
hexagonal lattice. The lattice constant and the width of the Gaussian
peaks are obtained from minimization of the functional (Eq(1)) with
respect to these
variables.
It can be seen directly, that the distribution function does not change
at the nematic-columnar coexistence.
The densities at coexistence, normalized by the density at close packing
($\eta_{cp}=\pi/2\sqrt{3}\sim0.907$),
are determined as follows: $\eta^{\rm nem}/\eta_{\rm cp}=0.36$ and
$\eta^{\rm col}/\eta_{\rm cp}=0.43$, independent of the polydispersity of the
system.

For standard deviations $\sigma <0.25$ we find a continuous
transition from the nematic to the smectic phase. The length distribution
does not change the continuous nature of the transition. This result
is in agreement with computer simulations by Stroobants$^7$, who found a
continuous nematic-smectic transition for a binary mixture of long and short
spherocylinders. This result is not at all obvious.
In principle one might expect that the nematic-smectic transition could
be accompanied by the separation of rods, similar to the isotropic-nematic
transition. In the latter case the longer rods are more abundant in the
nematic than in the isotropic phase$^{10,11,18,19}$. By analogy, we could
expect the length distribution to narrow at the nematic-smectic transition
and consequently to change the continuous nature of the transition.
The transition has been studied as follows. First we have calculated the
chemical potential of the smectic phase. It reads:
$$\eqalign{&\mu^{\rm sm}(L)=\ln{(g^{\rm sm}(L))}+\cr&{\int dz\rho^{\rm sm}(z)
\ln{(\lambda^3\rho^{\rm sm}(z))}\over{\int dz\rho^{\rm sm}(z)}}+
{{\int dz\rho^{\rm sm}(z)\Psi(v_0\bar\rho^{\rm sm}(z,L))}\over
{\int dz\rho^{\rm sm}(z)}}+\cr
&{{\int dz\int dz'\int dL' f_1(L+L',\vert z-z'\vert)\rho^{\rm sm}(z')
v_0\rho^{\rm sm}(z)g^{\rm sm}(L')\Psi'
(v_0\bar\rho^{\rm sm}(z',L'))}\over{\int dz\rho^{\rm sm}(z)}}.\cr}
\eqno(11)$$
where $f_1$ is defined in Eq(5).
Then we have equated nematic and smectic chemical potentials:
$$\mu^{\rm nem}(L)=\mu^{\rm sm}(L).\eqno(12)$$
It follows immediately, that for the known distribution function
$g^{\rm sm}(L)$, the distribution function $g^{\rm nem}(L)$ is
trivially determined at coexistence by Eq(8), i.e.,
$$g^{\rm nem}(L)=\exp{\left(\mu^{\rm sm}(L)-\mu^{\rm nem}_0\right)}\eqno(13)$$
The second equilibrium condition, i.e., the equality of pressure, combined
with the normalization condition $\int dLg^{\rm nem}(L)=1$ determines the
coexisting densities. We have found further that for all degrees
of polydispersity $\sigma$ the nematic-smectic
phase transition is continuous.

The same procedure has been applied to the columnar-smectic phase transition.
In this case we have assumed a given length distribution function in the
smectic
phase and from the equality of chemical potentials (Eqs(8,10,11))
we could determine the distribution function in the columnar phase
at coexistence with the smectic. This transition is first order,
so the average volume fraction $\eta$ jumps at the
transition. The distribution function in the smectic is again assumed
to be Gaussian. We characterize the distribution function in the columnar
by its mean value and standard deviation
in order to compare it to the distribution function in the smectic.
It turns out that the change in the mean length is negligible. The standard
deviation of the distributions is larger in the columnar than in the
smectic. This is the expected result, since the lamellar ordering in
the smectic favours a sharp distribution whereas the columnar structure
does not. For example, as seen in Fig.1, a smectic phase with a
polydispersity of $\sigma=0.15$ at a
packing-fraction of $\eta/\eta_{\rm cp}=0.45$ is in equilibrium with a
columnar structure of a polydispersity of $\sigma=0.17$ at a
packing-fraction of $\eta/\eta_{\rm cp}=0.50$.

The dashed lines shown in Fig.1 represent the instability of the smectic and
columnar phases with respect to perturbations to a hexagonal solid.
The phase characterized
by the density distribution $\rho^{i}({\bf r})$ is unstable with respect
to the perturbation $\delta\rho^{(i,f)}({\bf r})$ of the symmetry of the
$f$ phase if the following condition holds:
$$\int d{\bf r}\int d{\bf r}'\delta\rho^{(i,f)}({\bf r})\delta\rho^{(i,f)}
({\bf r}'){{\delta^2 F[\rho]}\over{\delta\rho({\bf r})\delta\rho({\bf r})}}
\Biggl\vert_{\rho({\bf r})=\rho^i({\bf r})}=0\eqno(14)$$
We have assumed that the perturbations can be expressed in the factorized form:
$$\delta\rho^{(i,f)}({\bf r})=\rho^i({\bf r})\delta\rho^f({\bf r}),\eqno(15)$$
where $\delta\rho^f({\bf r})$ describes the onset of the ordering
specific for the phase $f$ and absent in the phase $i$.
For analysis of the columnar solid bifurcation we take
$$\delta\rho^{f}({\bf r})=\cos{(k z)}\eqno(16)$$
while for the smectic solid bifurcations we assume
$$\delta\rho^{f}({\bf r})=\sum_{n=1}^3\cos({\bf k}_n {\bf r}_\bot),\eqno(17)$$
where $k_1=(1,1/\sqrt{3})k$, $k_2=(-1,1/\sqrt{3})k$ and $k_3=(0,2/\sqrt{3})k$
are the vectors spanning the first shell in the reciprocal space for the
regular hexagonal lattice.

For large polydispersity the columnar phase stabilizes with
respect to solid perturbations. We expect this, since in the system of
rods with continuous distribution of length, the particles
do not fit well into the 3D structure involving ordering
along the long axis of rods.
This transition is expected to be first order, with the distribution function
more strongly peaked in the solid. To quantify this in detail,
it is necessary to
examine the columnar-solid coexistence.
The smectic phase becomes slightly destabilized with respect to solid
perturbations if the polydispersity is increased. This can be understood as
an indication that order in the z-direction is least favoured, so that even
3-dimensional ordering is preferred.

To complete this study, we investigate a binary mixture of rods of two
different lengths. The aim is
to find out, under what conditions we have to expect any
form of demixing. Since a bidisperse
system is more likely to segregate into its components than a polydisperse
one, we consider this to be a stronger criterion than looking for
an instability of a polydisperse system.
We have already found, that no demixing
occurs at the nematic-smectic transition in the polydisperse system.
This result remains also valid in the binary mixture.

The functional given by Eq(1) for the
general case of a polydisperse system reduces to the binary mixture case
for
$$\rho({\bf r},L)=\rho_1({\bf r})\delta_{L,L_1}+\rho_2({\bf r})
\delta_{L,L_2},\eqno(18)$$
where $\delta_{i,j}$ is the Kronecker delta function.
Using this density distribution, we are able to calculate the free energy of
the nematic, smectic and columnar phase as well as the instabilities to
solid ordering as before. Fig.2 shows the results for different ratios
of the lengths of the rods. The phase-diagram is calculated at the
equivalence point, where the partial volume fractions of the two components
are the same. Although the results are not completely comparable
to those of Stroobants$^7$, who studied spherocylinders rather than cylinders,
qualitative agreement can still be seen easily. As in the study of
Stroobants, we observe the
destabilization of the smectic order
compared to the nematic, and stabilization
of the columnar order with an increasing length-ratio
of the two components.
The qualitative similarity of the phase diagrams for the polydisperse and
bidisperse systems is remarkable (compare Fig.1 and Fig.2).
This is a quite unexpected result and indeed may have interesting
implications for the future study of such
systems.

We now consider the possibility of demixing
within the smectically ordered system. One might, for example, expect that
alternating layers of different widths would be formed in the bidisperse
system. We look only for a separation
of the complete system into one phase consisting mainly of long rods
and another phase consisting mainly of short rods, since
it can be shown,
that the free energy of such a system is comparable to
that of smectics with alternating layers.
In order to study the stability we check the
partial derivatives
$$M_{ij}={{\partial^2 F[\rho]}\over{\partial \rho_i\partial \rho_j}},
\eqno(19)$$
where $\rho_i$ are the partial densities of the two components.
It turns out that the matrix ($M_{ij}$) remains positive definite for all
length-ratios, compositions and packing-fractions, for which the smectic phase
is stable. We also check for the possible coexistence
between two smectic phases with widely varying composition,
by equating the chemical potentials of both components as well as the
pressure. Again we find no such coexistence in the smectic phase.
This leads to the conclusion that demixed smectics or smectics with
varying lamella-widths are not found in binary systems under the present
conditions. It follows
that demixed smectics are unlikely to occur in multi-component systems.
Demixing may well occur in the solid, however.

A physical realization of our model is a dense
polydisperse suspension of
elongated colloidal particles stabilized against irreversible aggregation
by surface-grafted polymer layers.
In a good solvent
the interactions between such colloidal particles can be approximated by
hard core interactions$^{20,21}$.

In general, however, macromolecules in
solutions interact via van der Waals forces, and if they carry surface
charges, also via screened Coulombic repulsion$^{20,22}$.
The influence of these complex interactions on the phase diagram
is not known in general.

Different problems arise when the macromolecules are not rigid but
semi\-flexible$^{23,24}$. It should be noted as well that in self assembling
systems, when the length distribution is given by the thermodynamic
conditions in a given phase$^{19,25}$, the equality of
monomer
chemical potential is sufficient to set the coexistence conditions
between two phases.

Summarizing; by means of density-functional theory, we have studied in
detail the influence of polydispersity
on phase equilibria in oriented rodlike macromolecular systems.
For a Gaussian distribution of rods length we find
three polydispersity regimes. For $0.15>\sigma\ge 0$ we have the following
sequence of phase transitions nematic-smectic-solid; for $0.25>\sigma>0.15$
we find the nematic-smectic-columnar and solid phases; for
$0.27>\sigma>0.25$ we find direct nematic-columnar and columnar-solid
transitions
and for $\sigma>0.27$ there are only nematic and columnar phases.
We also find that polydispersity does not affect the
continuous nature of the nematic-smectic phase transition. We do not find
any evidence for demixed smectic phases.
Our novel method for the study of polydispersity
 can be easily applied also to
the isotropic-nematic phase transition$^{11}$.

\centerline{\bf Acknowledgements}
We thank Dr.A.R.Denton and Prof.H.L\"owen for helpful discussions.
This work was supported in part by the KBN and {\it "Stiftung f\"ur
deutsch-polnische Zusammenarbeit''} grants. Financial support by the
{\it "Fonts der chemischen Industrie''} is also gratefully acknowledged.

\centerline{\bf References}
\item{1.} Feughelman et al, Nature {\bf 175}, (1955);
F.Livolant, A.M.Levelut, J.Doucet and J.P.Benoit, Nature {\bf 339},
724 (1986);
F.Livolant, J.Mol.Biol. {\bf 218}, 165 (1991); F.Livolant Physica A
{\bf 176}, 117 (1991);
D.Durand, J.Doucet and F.Livolant, J.Phys II {\bf 2}, 1769 (1992).
\item{2.} J.Watanabe and Y.Takashina, Macromolecules {\bf 24}, 3423 (1991).
\item{3.} F.Livolant and Y.Bouligand, J.Physique {\bf 48}, 1813 (1986).
\item{4.} J.M.Rodriguez-Parada, R.Duran and G.Wegner, Macromolecules
{\bf 22}, 2507 \hfill\break(1989); S.Schwiegk, T.Vahlenkamp, Y.Xu and G.Wegner,
Macromolecules {\bf 25}, 2513 (1992).
\item{5.} R.S.Fuller, J.M.Kaguni and A.Kornberg, Proc.Natl.Acad.Sci. {\bf 78},
7370, (1981); I.Baeza et al, Biochemistry {\bf 26}, 1387 (1987);
J.L.Sikorav and G.M.Church, J.Mol.Biol. {\bf 222}, 1085 (1991).
\item{6.} R.Ho\l yst and A.Poniewierski, Mol.Phys. {\bf 71}, 561 (1990);
J.A.C.Veerman and D.Frenkel, Phys.Rev.A {\bf 43}, 4334 (1991).
\item{7.} A.Stroobants, Phys.Rev.Lett. {\bf 69}, 2388 (1992);
R.P.Sear and G.Jackson,\hfill\break J.Chem.Phys. {\bf 102}, 2622 (1995).
\item{8.} R.Evans, Adv.Phys. {\bf 28}, 143 (1979);
H. L\"owen, Phys.Rep. {\bf 237}, 251 (1994).
\item{9.} R.Ho\l yst and A.Poniewierski, Phys.Rev.A {\bf 39}, 2742 (1989);
H.Xu, H.N.W. Lekkerkerker and M.Baus, Europhys.Lett. {\bf 17}, 163 (1992).
\item{10.} T.J.Sluckin, Liq.Cryst. {\bf 6}, 111 (1989).
\item{11.} Z.Y.Chen, Phys.Rev. E {\bf 50}, 2849 (1994).
\item{12.} P.G. de Gennes and J.Prost, {\it Physics of Liquid Crystals},
p. 551 Oxford Scientific Publications, 1993.
\item{13.} A.R. Denton and N.W. Ashcroft, Phys.Rev A {\bf 39}, 4701 (1989);
{\it ibid}, {\bf 42}, 7312 (1990).
\item{14.} D.A.Young and B.J.Alder, J.Chem.Phys. {\bf 60}, 1254 (1974).
\item{15.} A.Kyrlidis and R.A.Brown, Phys.Rev. E {\bf 47}, 427 (1993).
\item{16.} R.Ohnesorge, H.L\"owen and H.Wagner, Europhys.Lett. {\bf 22},
245 (1993).
\item{17.} H.L\"owen, private communication.
\item{18.} R.Diebleck and H.N.W.Lekkerkerker, J.Phys.Lett. (Paris), {\bf 41},
L351 (1980).
\item{19.} W.E.McMullen, W.M. Gelbart and A. Ben-Shaul, J.Chem.Phys. {\bf 82},
5616 (1985).
\item{20.} P.Pusey in {\it Liquids, Freezing and Glass Transitions} edited
by J.P.Hansen, \hfill\break D.Levesque and J.Zinn-Justin (North Holland,
Amsterdam 1991);
G.J.Vroege and H.N.W. Lekkerkerker, Rep.Prog.Phys. {\bf 55}, 1241 (1992).
\item{21} J.Prost and F.Rondelez, Nature (Suppl) {\bf 350}, 11 (1991).
\item{22.} J.N. Israelachvili, {\it Intermolecular and surface forces},
Academic Press (1985).
\item{23.} D.R.Nelson in {\it Observations, Predictions and Simulations of
Phase Transitions in Complex Fluids} ed by M.Baus et al, p.293,
Kluwer Academic Publisher (1995).
\item{24.} J.V.Selinger  and R.F.Bruinsma, Phys.Rev. A {\bf 43}, 2922 (1991).
\item{25.} M.P.Taylor and J.Herzfeld, Langmuir {\bf 6}, 911 (1990);
Phys.Rev. A {\bf 43}, 1892 (1991).
\vfill\eject
\centerline{\bf Figure Captions}
\item{Fig.1} Phase diagram for a polydisperse system of parallel rods
interacting via hard core repulsive forces. The polydispersity of the rod
lengths is modelled by a Gaussian distribution of standard deviation $\sigma$.
Squares denote coexistence of phases, triangles a second order phase
transition, and crosses represent instabilities.
\item{Fig.2} Phase diagram for a two component system of length ratio
$L_1/L_2$. Note the remarkable qualitative similarity to Fig.1.
\vfill\eject
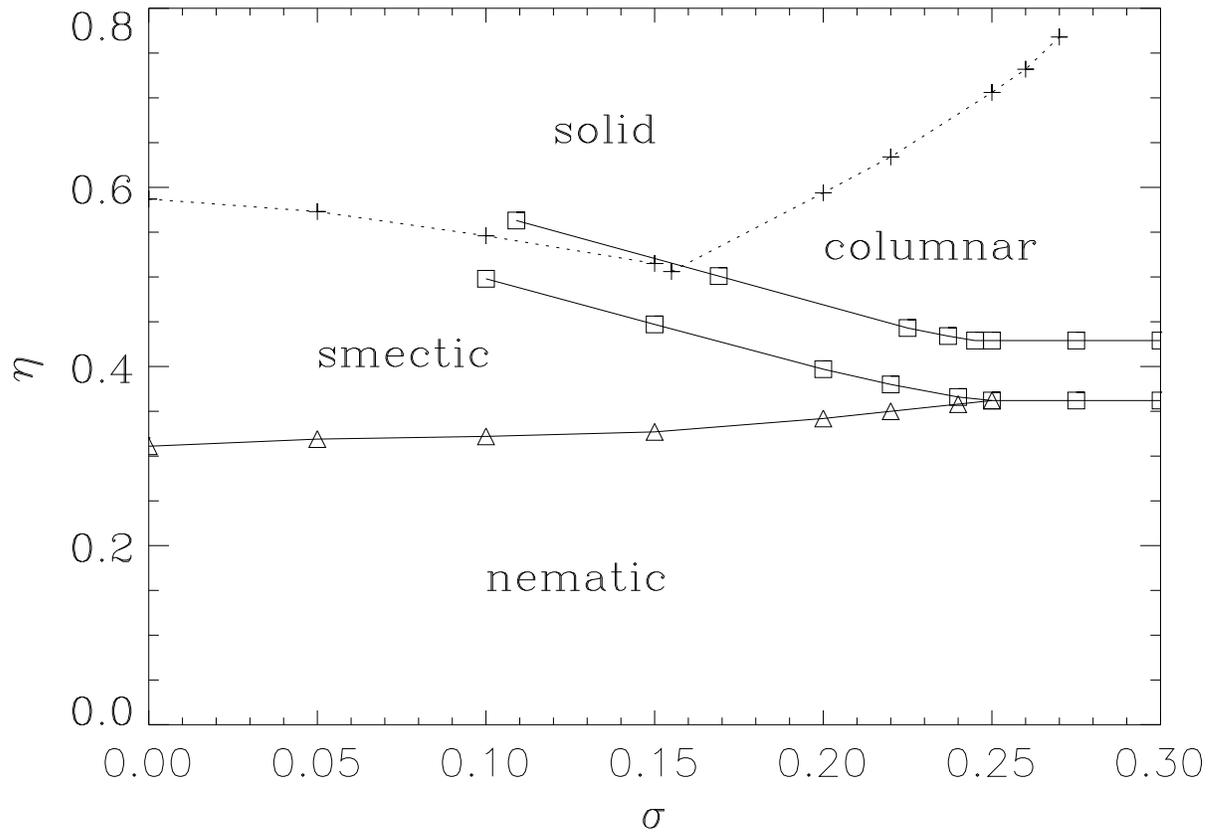
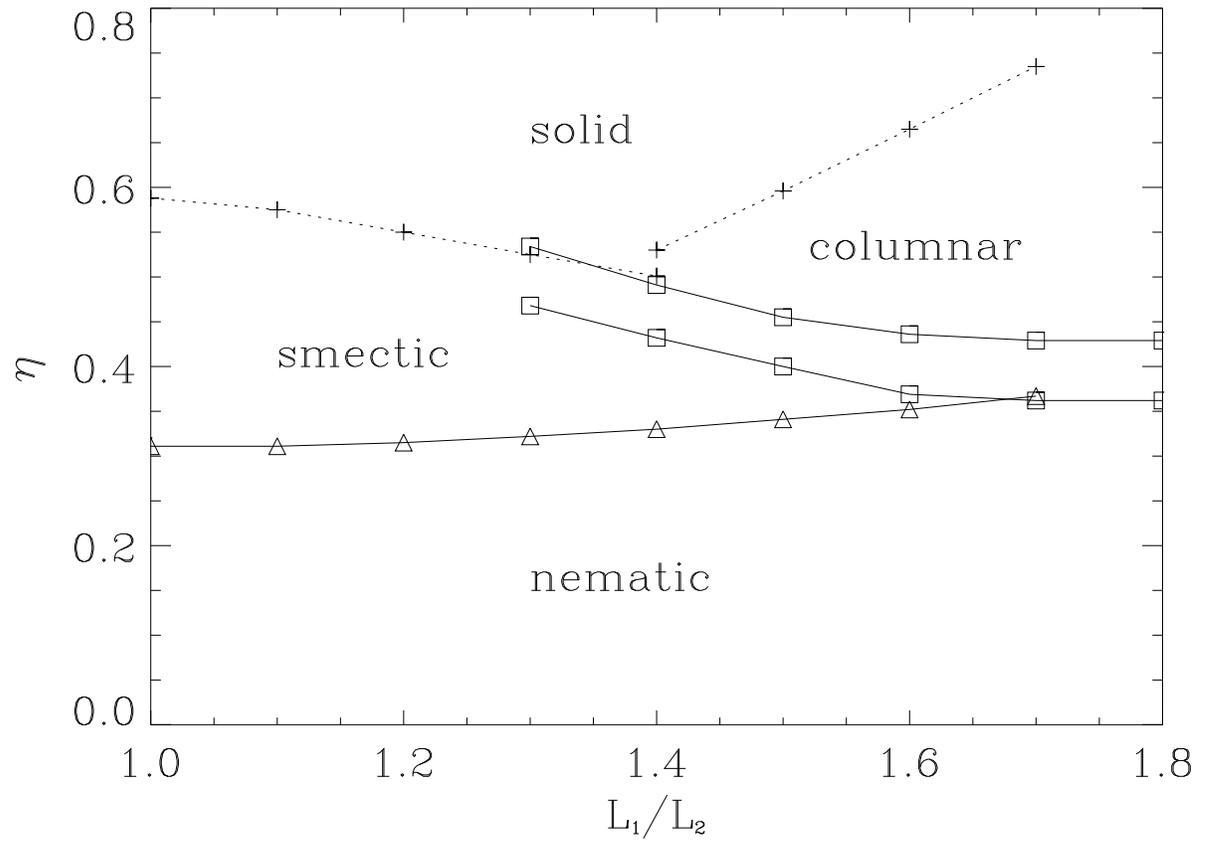
\end